\documentclass[12pt]{article}

\begin{document}

\title{Four-Index Energy-Momentum Tensors for Gravitation and Matter\footnote{Published in: ''Z.Zakir (2003) \textit{Structure of Space-Time and Matter}, CTPA, Tashkent.'' }}
\author{Zahid Zakir\thanks{E-mail: zahid@in.edu.uz}\\Centre for Theoretical Physics and Astrophysics,\\ P.O.Box 4412, Tashkent 700000 Uzbekistan}
\date{May 4, 1999;\\
Revised October 17, 2003. }
\maketitle

\begin{abstract}
The 4-index energy-momentum tensors for gravitation and matter are analyzed on
the basis of new equations for the gravitational field with the Riemann
tensor. Some properties of the such defined gravitational energy are discussed.
\end{abstract}

\section{Introduction}

In general relativity a true and covariant characteristics of the
gravitational field is the Riemann curvature tensor $R_{iklm}$, but the field
equations contain only the Ricci tensor, vanishing in the vacuum. The Weyl
tensor $C_{iklm}$ which is a nonvanishing in the vacuum pure 4-index part of
$R_{iklm}$, disappear at 2-index contraction \cite{Za}. This fact was leading
to the problems with the definition of the energy-momentum for the
gravitational field.

In the paper \cite{Za} a new generalized version of the Einstein equations
with the Riemann tensor has been formulated. It has been shown that the
4-index energy-momentum tensors for gravitation and matter can be constructed.
In the present paper the structure of new gravitational equations and
properties of 4-index energy-momentum tensors will be discussed.

\section{Four-index equations for gravitation}

We started from the standard Einstein-Gilbert action for the gravitational field:%

\begin{eqnarray}
S  &  =&\frac{1}{2}\int d\Omega\sqrt{-g}\left(  -\frac{1}{\kappa}R+L\right)
=\\
&  =&-\frac{1}{2}\int d\Omega\sqrt{-g}\left[  \frac{1}{2\kappa}(g^{il}%
g^{km}-g^{im}g^{kl})R_{iklm}-L\right]  ,
\end{eqnarray}
where $\kappa=8\pi k/c^{4}$. We perform the variational procedure so that
$R_{iklm}$ preserves its 4-index form. The result of the such variation is
\cite{Za}:%

\begin{equation}
\delta S_{g}=-\frac{1}{2}\int d\Omega\sqrt{-g}\delta g^{km}g^{il}%
[G_{iklm}-T_{iklm}]=0.
\end{equation}
Here new 4-index tensors are defined as ($d=4$):%

\begin{eqnarray}
G_{iklm}  &  =&\frac{1}{\kappa}\left[  R_{iklm}-\frac{1}{6}(g_{il}g_{km}%
-g_{im}g_{kl})R\right]  ,\\
T_{iklm}  &  =&V_{iklm}+\frac{1}{2}(g_{km}T_{il}-g_{kl}T_{im}+g_{il}%
T_{km}-g_{im}T_{kl})-\nonumber\\
& &  -\frac{1}{6}(g_{il}g_{km}-g_{im}g_{kl})T,
\end{eqnarray}
where $V_{iklm}$, having the property $g^{il}V_{iklm}=0$ and which does not
vanish in the vacuum around the source, can be identified by the required
energy-momentum density tensor for gravitational field. The field equations in
the general case are:%

\begin{equation}
G_{iklm}=T_{iklm}.
\end{equation}

The tensors $G_{iklm}$ and $T_{iklm}$ have the symmetry properties as the
Riemann tensor and therefore we have 20 equations. The tensor $G_{iklm}$ is a
function of the metric tensor $g_{ik}$ which has 6 independent components. The
tensor $T_{iklm}^{(m)}$ is combined from the ordinary energy-momentum tensor
of the matter $T_{ik}$ and it has 4 independent functions (the energy density
$\epsilon$ and 3 components of the velocity). These 10 functions are solutions
of 10 Einstein equations (6 for independent components of the metric and 4 for
independent components of $T_{ik}$). The new term $V_{iklm}$ has 10
independent components.

So, we have 20 equations for 20 independent functions. If we take solutions of
the Einstein equations for some metric and $T_{ik}$, then we have an
additional 10 equations for 10 components of $V_{iklm}$. This means that the
solutions of the Einstein equations exactly define all components of
$V_{iklm}$ and we can find $V_{iklm}$ for some standard metric. But if we have
some model of the vacuum and calculate $V_{iklm}$ in this model, then we have
10 equations for 10 unknown components of the metric $g_{ik}$ and $T_{ik}.$

\section{The energy-momentum conservation for the system of gravitation and matter}

The Riemann tensor can be represented as:
\begin{eqnarray}
R_{iklm}  &  =&C_{iklm}+\frac{1}{2}(g_{km}R_{il}-g_{kl}R_{im}+g_{il}%
R_{km}-g_{im}R_{kl})-\\
&  &-\frac{1}{6}(g_{il}g_{km}-g_{im}g_{kl})R
\end{eqnarray}
where $C_{iklm}$ is the Weyl tensor with zero 2-index contraction
$g^{il}C_{iklm}=0$. In the vacuum $T_{ik}=T=0,$ $R_{il}=R=0$, and we have:%

\begin{equation}
\frac{1}{\kappa}C_{iklm}=V_{iklm}%
\end{equation}

The covariant derivatives of these 4-index tensors are:
\begin{eqnarray}
G_{.klm;i}^{i}  &  =&\frac{1}{\kappa}\left[  R_{.klm;i}^{i}-\frac{1}{6}%
(g_{km}R_{,l}-g_{kl}R_{,m})\right]  =\nonumber\\
&  =&T_{km;l}-T_{kl;m}-\frac{1}{3}(g_{km}T_{,l}-g_{kl}T_{,m}),\\
T_{klm;j}^{j(m)}  &  =&\frac{1}{2}\left[  T_{km;l}-T_{kl;m}-\frac{1}{3}%
(g_{km}T_{;l}-g_{kl}T_{;m})\right]  =\frac{1}{2}G_{.klm;i}^{i}.
\end{eqnarray}
Then we obtain the relationship:%

\begin{equation}
V_{klm;j}^{j}=G_{klm;j}^{j}-T_{klm;j}^{j(m)}=\frac{1}{2}G_{.klm;i}^{i}.
\end{equation}
and, therefore,
\begin{equation}
V_{klm;j}^{j}=T_{klm;j}^{j(m)}%
\end{equation}

In the vacuum, therefore, there are local conservation laws:%

\begin{equation}
G_{\cdot klm;j}^{j}=V_{klm;j}^{j}=0.
\end{equation}

The integral energy-momentum tensors for matter and gravity can be defined as:%

\begin{eqnarray}
P_{ikl}^{(m)}  &  =&\int dS_{n}\sqrt{-g}T_{.ikl}^{n(m)},\\
V_{ikl}  &  =&\int dS_{n}\sqrt{-g}V_{.ikl}^{n}.
\end{eqnarray}

In the hypersurface $x^{0}=const$ we have:%

\begin{eqnarray}
P_{ikl}^{(m)}  &  =&\int d^{3}x\sqrt{-g}T_{.ikl}^{0(m)},\\
V_{ikl}  &  =&\int d^{3}x\sqrt{-g}V_{.ikl}^{0}.
\end{eqnarray}

We can construct the angular momentum tensor for the gravitational field as:%

\begin{equation}
M^{ilkm}=\int dS_{n}\sqrt{-g}(x^{m}V^{nikl}-x^{i}V^{nmkl}).
\end{equation}
This 4-index tensor can be interpreted as \textit{the} \textit{angular
momentum tensor} for the gravitational field.

\section{Comparison with pseudotensor and Hamiltonian approaches}

The pseudotensor $t_{ik}$, for example, in the Landau-Lifshitz version, has
been defined as a part of the Einstein tensor:%

\begin{equation}
G^{ik}=\frac{1}{(-g)}\frac{\partial\psi^{ikl}}{\partial x^{l}}-t^{ik}.
\end{equation}
The such separation leads to the conservation of the sum of $t^{ik}$ and
$T^{ik}$:%

\begin{equation}
\int dS_{k}(-g)(T^{ik}+t^{ik})=\int dS_{k}\frac{\partial\psi^{ikl}}{\partial
x^{l}},
\end{equation}
and the right hand term then can be interpreted as a total energy of the system.

Thus, if we want to work with a localizable and tensor form of the
gravitational energy-momentum in the vacuum, we can take the Weyl tensor,
which has zero 2-index contraction. If we want to work with non-zero 2-index
form of the gravitational energy-momentum, we take the pseudotensors or the
Hamiltonians, which are non-localizable and non-covariant.

\subsection{The geodesic deviation and measurements of the gravitational energy}

The equation of geodesic deviation:%

\begin{equation}
\frac{D^{2}\eta^{i}}{ds^{2}}=R_{klm}^{i}u^{k}u^{l}\eta^{m}%
\end{equation}
we rewrite in vacuum as:%

\begin{equation}
\frac{D^{2}\eta^{i}}{ds^{2}}=C_{klm}^{i}u^{k}u^{l}\eta^{m}=\kappa V_{klm}%
^{i}u^{k}u^{l}\eta^{m}.
\end{equation}

Therefore, we conclude that the measurements of the geodesic deviations are
exactly the measurements of the 4-index energy-momentum tensor of
gravitational field $V_{klm}^{i}$.

In \cite{Za 2} the energies for standard metrics and some consequences of the
proposed treatment of the gravitational energy will be considered.

\end{document}